\documentclass[a4paper,11pt]{article}

\usepackage{hyperref}
\hypersetup{
colorlinks=true,
linkcolor=magenta,
citecolor=blue,
urlcolor=black,
bookmarksnumbered=true,
bookmarkstype=toc,
bookmarksopen=false,
pdftitle={Thermal Correlators from Rindler-AdS2/CFT1},
pdfauthor={Satoshi Ohya},
pdfsubject={conformal field theory},
pdfkeywords={two-point function; AdS/CFT correspondence}}

\usepackage[margin=1in]{geometry}

\usepackage[adobe-utopia,cal=cmcal]{mathdesign}
\usepackage[T1]{fontenc}

\usepackage{xcolor}

\usepackage{amsmath,amsbsy,bm}

\usepackage{booktabs,multirow}
\usepackage[font=small,labelfont=bf]{caption}

\usepackage{cite}

\title{
\Large\bfseries\boldmath Thermal Correlators from Rindler-AdS$_{2}$/CFT$_{1}$\footnote{Contribution to the proceedings of the 5th CST-MISC Joint Symposium on Particle Physics, Tokyo, Japan, October 24-25, 2015.}
}
\author{
Satoshi Ohya\\[2ex]
\textit{\small Institute of Quantum Science, Nihon University}\\
\textit{\small Kanda-Surugadai 1-8-14, Chiyoda, Tokyo 101-8308, Japan}\\[1ex]
\texttt{\small E-mail:\,\href{mailto:ohya@phys.cst.nihon-u.ac.jp}{ohya@phys.cst.nihon-u.ac.jp}}
}
\date{\small (Dated: \today)}

\begin{document}
\maketitle

\begin{abstract}
In this paper we study one-dimensional conformal field theory at finite temperature dual to the two-dimensional anti-de Sitter spacetime in the Rindler coordinates.
We show that conformal symmetry for thermal two-point functions manifests itself in a form of recurrence relations in the complex frequency space.
It is discussed that all the real-time two-point functions are given by solutions to the recurrence relations.
\end{abstract}

\section{Introduction} \label{sec:1}
The purpose of this paper is to report a simple Lie algebraic approach to momentum-space two-point functions of finite-temperature conformal field theory initiated in Refs.~\cite{Ohya:2013xva,Ohya:2013vba}.
For the sake of simplicity, in this paper we would like to focus on frequency-space thermal two-point functions for a scalar primary operator of one-dimensional conformal field theory (CFT$_{1}$).
We shall show that conformal symmetry for thermal two-point functions $G_{\Delta}(\omega)$ in frequency space manifests itself in the form of linear functional relations (or \textit{recurrence relations} in the complex $\omega$-plane)
\begin{align}
G_{\Delta}(\omega)
&= 	\frac{- 1 + \Delta \pm \tfrac{i\omega}{2\pi T}}{- \Delta \pm \tfrac{i\omega}{2\pi T}}
	G_{\Delta}(\omega \pm i2\pi T), \label{eq:1}
\end{align}
where $\Delta$ is the scaling dimension of scalar primary operator and $T$ is the temperature.
All the real-time two-point functions are turned out to satisfy these recurrence relations.
To derive Eq.~\eqref{eq:1}, we utilize the AdS/CFT correspondence and consider the two-dimensional anti-de Sitter spacetime in the Rindler coordinates (Rindler-AdS$_{2}$).
The Rindler-AdS$_{2}$ (\textit{aka} the AdS$_{2}$ black hole \cite{Spradlin:1999bn}) is just a portion of AdS$_{2}$ where the time-translation Killing vector generates the noncompact Lorentz group $SO(1,1)$; see Table \ref{tab:1}.
In an appropriate parameterization such an AdS$_{2}$ portion can be described by the following black-hole-like metric:
\begin{align}
ds_{\text{Rindler-AdS}_{2}}^{2}
&= 	- \left(\frac{r^{2}}{\ell^{2}} - 1\right)dt^{2}
	+ \frac{dr^{2}}{r^{2}/\ell^{2} - 1},
	\quad
	r \in (\ell,\infty), \label{eq:2}
\end{align}
where $r=\infty$ corresponds to the AdS$_{2}$ boundary while $r=\ell$ the Rindler horizon with $\ell$ being the AdS$_{2}$ radius.
In contrast to the well-known global and Poincar\'{e} coordinate patches, the Rindler-AdS$_{2}$ becomes thermal thanks to the $SO(1,1)$ time-translation group and is known to be dual to a finite-temperature CFT$_{1}$ with nonzero temperature $T = 1/(2\pi\ell)$.\footnote{Thermal aspects of Rindler-AdS have also been studied in \cite{Czech:2012be,Parikh:2012kg}.}
The goal of this paper is to show that, with the aid of the real-time prescription of the AdS/CFT correspondence \cite{Iqbal:2008by,Iqbal:2009fd}, the recurrence relations \eqref{eq:1} just follow from the $SO(2,1)$ isometry of Rindler-AdS$_{2}$.
To see this, it is convenient to introduce a new spatial coordinate $x$ via
\begin{align}
r = \ell\coth(x/\ell), \quad x \in (0,\infty), \label{eq:3}
\end{align}
where $x=0$ corresponds to the AdS$_{2}$ boundary while $x=\infty$ the Rindler horizon.
It is easy to check that in this new coordinate system $(t,x)$ the metric becomes conformally flat:
\begin{align}
ds_{\text{Rindler-AdS}_{2}}^{2} = \frac{-dt^{2}+dx^{2}}{\sinh^{2}(x/\ell)}. \label{eq:4}
\end{align}
In the following we shall analyze a finite-temperature CFT$_{1}$ residing on the boundary $x=0$.

The rest of the paper is organized as follows.
In Section \ref{sec:2} we first briefly discuss unitary representations of the Lie algebra $\mathfrak{so}(2,1)$ in the basis in which the $SO(1,1)$ generator becomes diagonal and then introduce a coordinate realization of $SO(2,1)$ generators which are given by the Killing vectors of Rindler-AdS$_{2}$.
We then derive the recurrence relations \eqref{eq:1} for frequency-space two-point functions in Section \ref{sec:3}.
We shall see that two-point Wightman functions and advanced/retarded two-point functions are obtained as ``minimal'' solutions of Eq.~\eqref{eq:1}.

In the rest of the paper we will work in the units $\ell = 1$ (i.e., $2\pi T = 1$).

\begin{table}[t]
\begin{center}
\begin{tabular}{l c c c c}
\toprule
\multirow{2}{*}{coordinate patch} 	& \multicolumn{2}{c}{time-translation group} 	& \multicolumn{2}{c}{frequency spectrum} \\
			& Lorentzian 	& Euclidean 			& Lorentzian 	& Euclidean \\
\midrule
global 		& $SO(2)$ 	& $SO(1,1)$ 			& discrete 	& continuous \\
Poincar\'{e} 	& $E(1)$ 		& $E(1)$ 				& continuous 	& continuous \\
Rindler 		& $SO(1,1)$ 	& $SO(2)$ 			& continuous 	& discrete \\
\bottomrule
\end{tabular}
\caption{Three distinct AdS$_{2}$ coordinate patches. The time-translation Killing vectors in the global, Poincar\'{e} and Rindler coordinates in Lorentzian signature generate the one-parameter subgroups $SO(2), E(1), SO(1,1) \subset SO(2,1)$, respectively. It can be understood from the following that the Rindler-AdS$_{2}$ becomes thermal: i) the noncompact Lorentz group $SO(1,1)$ becomes the compact rotation group $SO(2)$ under the Wick rotation; ii) the spectrum of compact $SO(2)$ is always discretized; and iii) discretized frequencies in Euclidean signature are nothing but the Matsubara frequencies in finite-temperature field theory.}
\label{tab:1}
\end{center}
\end{table}

\section{\boldmath UIR of \texorpdfstring{$SO(2,1)$}{SO(2,1)} in the \texorpdfstring{$SO(1,1)$}{SO(1,1)} diagonal basis} \label{sec:2}
Let us start with the Lie algebra $\mathfrak{so}(2,1)$ of the one-dimensional conformal group $SO(2,1)$, which is spanned by the three self-adjoint generators $\{J_{1}, J_{2}, J_{3}\}$ satisfying the commutation relations
\begin{align}
[J^{1},J^{2}] = +iJ^{3}, \quad
[J^{2},J^{3}] = -iJ^{1}, \quad
[J^{3},J^{1}] = -iJ^{2}. \label{eq:5}
\end{align}
Note that $J^{3}$ generates the compact rotation group $SO(2)$, whereas $J^{1}$ and $J^{2}$ generate the noncompact Lorentz group $SO(1,1)$.
We are interested in Unitary Irreducible Representations (UIRs) of $SO(2,1)$ in the basis in which the $SO(1,1)$ generator becomes diagonal, because, as mentioned in Section \ref{sec:1}, in the Rindler-AdS$_{2}$ the time-translation Killing vector generates the noncompact subgroup $SO(1,1) \subset SO(2,1)$.
In order to study the UIRs in the $SO(1,1)$ diagonal basis, we introduce the following \textit{hermitian} linear combinations:
\begin{align}
J^{\pm} := J^{2} \pm J^{3}, \label{eq:6}
\end{align}
which satisfy the commutation relations
\begin{align}
[J^{1}, J^{\pm}] = \pm iJ^{\pm}, \quad
[J^{+}, J^{-}] = 2iJ^{1}. \label{eq:7}
\end{align}
The quadratic Casimir $C$ of the Lie algebra $\mathfrak{so}(2,1)$ is given by
\begin{align}
C = -(J^{1})^{2}-(J^{2})^{2}+(J^{3})^{2} = -J^{1}(J^{1} \pm i) - J^{\mp}J^{\pm}, \label{eq:8}
\end{align}
which commutes with all the generators, $[C, J^{a}] = 0$ ($a=1,2,3$), as it should.
Let $|\Delta,\omega\rangle$ be a simultaneous eigenstate of the Casimir operator $C$ and the $SO(1,1)$ generator $J^{1}$ that satisfy the eigenvalue equations
\begin{subequations}
\begin{align}
C|\Delta,\omega\rangle
&= 	\Delta(\Delta-1)|\Delta,\omega\rangle, \label{eq:9a}\\
J^{1}|\Delta,\omega\rangle
&= 	\omega|\Delta,\omega\rangle. \label{eq:9b}
\end{align}
\end{subequations}
It then follows from the commutation relations $[J^{1}, J^{\pm}] = \pm iJ^{\pm}$ that the states $J^{\pm}|\Delta,\omega\rangle$ satisfy the relations $J^{1}J^{\pm}|\Delta, \omega\rangle = ([J^{1}, J^{\pm}] + J^{\pm}J^{1})|\Delta, \omega\rangle = (\omega \pm i)J^{\pm}|\Delta, \omega\rangle$, which imply that $J^{\pm}$ raises and lowers the eigenvalue $\omega$ by $\pm i$:
\begin{align}
J^{\pm}|\Delta, \omega\rangle \propto |\Delta, \omega \pm i\rangle. \label{eq:10}
\end{align}

Let us now turn to the Rindler-AdS$_{2}$ problem.
In the Rindler coordinates the Killing vectors of AdS$_{2}$ are turned out to be given by the following first-order differential operators:
\begin{subequations}
\begin{align}
J^{1}
&= 	i\partial_{t}, \label{eq:11a}\\
J^{\pm}
&= 	\mathrm{e}^{\pm t}\left[\sinh x(i\partial_{x}) \pm \cosh x(i\partial_{t})\right], \label{eq:11b}
\end{align}
\end{subequations}
which indeed satisfy the commutation relations \eqref{eq:7} and are (formally) self-adjoint with respect to the integration measure $\sqrt{|g|}d^{2}x = dtdx/\sinh^{2}x$.
A straightforward calculation shows that the quadratic Casimir just gives the d'Alembertian $\Box = (1/\sqrt{|g|})\partial_{\mu}\sqrt{|g|}g^{\mu\nu}\partial_{\nu}$ on the Rindler-AdS$_{2}$
\begin{align}
C = \sinh^{2}x\left(- \partial_{t}^{2} + \partial_{x}^{2}\right) = \Box. \label{eq:12}
\end{align}
The eigenvalue equations \eqref{eq:9a} and \eqref{eq:9b} then reduce to the following differential equations:
\begin{subequations}
\begin{align}
i\partial_{t}\Phi_{\Delta,\omega}(t,x)
&= 	\omega\Phi_{\Delta,\omega}(t,x), \label{eq:13a}\\
\left(-\partial_{x}^{2} + \frac{\Delta(\Delta-1)}{\sinh^{2}x}\right)\Phi_{\Delta,\omega}(t,x)
&= 	\omega^{2}\Phi_{\Delta,\omega}(t,x). \label{eq:13b}
\end{align}
\end{subequations}
Notice that Eq.~\eqref{eq:13b} is nothing but the Klein-Gordon equation $(\Box-m^{2})\Phi_{\Delta,\omega}=0$ for a scalar field of definite frequency, $\Phi_{\Delta,\omega}(t,x) = \mathrm{e}^{-i\omega t}\Phi_{\Delta,\omega}(x)$, where the bulk mass $m^{2}$ and the scaling dimension $\Delta$ of dual CFT$_{1}$ operator are related as $\Delta(\Delta-1) = m^{2}$.

\section{Thermal two-point functions} \label{sec:3}
A finite-temperature CFT$_{1}$ dual to the Rindler-AdS$_{2}$ lives on the boundary $x=0$.
To analyze this, let us consider the following asymptotic near-boundary behaviors of Killing vectors:
\begin{subequations}
\begin{align}
J_{\text{near}}^{1}
&:= 	\lim_{x\to0}J^{1}
= 	i\partial_{t}, \label{eq:14a}\\
J_{\text{near}}^{\pm}
&:= 	\lim_{x\to0}J^{\pm}
= 	\mathrm{e}^{\pm t}\left(ix\partial_{x} \pm i\partial_{t}\right), \label{eq:14b}
\end{align}
\end{subequations}
which of course satisfy the commutation relations \eqref{eq:7}.
The near-boundary quadratic Casimir $C_{\text{near}} = -J_{\text{near}}^{1}(J_{\text{near}}^{1} \pm i) - J_{\text{near}}^{\mp}J_{\text{near}}^{\pm}$ takes the following simple form:
\begin{align}
C_{\text{near}} = x^{2}\partial_{x}^{2}. \label{eq:15}
\end{align}
The eigenvalue equations \eqref{eq:9a} and \eqref{eq:9b} near the boundary then become
\begin{subequations}
\begin{align}
i\partial_{t}\Phi^{\text{near}}_{\Delta,\omega}(t,x)
&= 	\omega\Phi^{\text{near}}_{\Delta,\omega}(t,x), \label{eq:16a}\\
\left(-\partial_{x}^{2} + \frac{\Delta(\Delta-1)}{x^{2}}\right)\Phi^{\text{near}}_{\Delta,\omega}(t,x)
&= 	0, \label{eq:16b}
\end{align}
\end{subequations}
which are easily solved with the result
\begin{align}
\Phi^{\text{near}}_{\Delta,\omega}(t,x)
&= 	A_{\Delta}(\omega)x^{\Delta}\mathrm{e}^{-i\omega t}
	+ B_{\Delta}(\omega)x^{1-\Delta}\mathrm{e}^{-i\omega t}, \label{eq:17}
\end{align}
where $A_{\Delta}(\omega)$ and $B_{\Delta}(\omega)$ are integration constants which may depend on $\Delta$ and $\omega$.
Substituting this into the ladder equations $J_{\text{near}}^{\pm}\Phi^{\text{near}}_{\Delta,\omega} \propto \Phi^{\text{near}}_{\Delta,\omega \pm i}$ we get
\begin{align}
&(i\Delta \pm \omega)A_{\Delta}(\omega)x^{\Delta}\mathrm{e}^{-i(\omega \pm i)t}
+ (i(1-\Delta) \pm \omega)B_{\Delta}(\omega)x^{1-\Delta}\mathrm{e}^{-i(\omega \pm i)t} \nonumber\\
&\propto
	\phantom{i\Delta}
	A_{\Delta}(\omega \pm i)x^{\Delta}\mathrm{e}^{-i(\omega \pm i)t}
	+
	\phantom{i(1-\Delta)\omega}
	B_{\Delta}(\omega \pm i)x^{1-\Delta}\mathrm{e}^{-i(\omega \pm i)t}, \label{eq:18}
\end{align}
which imply the following relations:
\begin{subequations}
\begin{align}
(i\Delta \pm \omega)A_{\Delta}(\omega)
&\propto 	A_{\Delta}(\omega \pm i), \label{eq:19a}\\
(i(1-\Delta) \pm \omega)B_{\Delta}(\omega)
&\propto 	B_{\Delta}(\omega \pm i). \label{eq:19b}
\end{align}
\end{subequations}
Now we are in a position to derive thermal two-point functions for a scalar primary operator of scaling dimension $\Delta$.
According to the real-time prescription of the AdS/CFT correspondence, the frequency-space two-point function for a scalar primary operator $\mathcal{O}_{\Delta}$ of scaling dimension $\Delta$ in the dual CFT$_{1}$ is given by the ratio \cite{Iqbal:2008by,Iqbal:2009fd}
\begin{align}
G_{\Delta}(\omega)
&= 	(2\Delta - 1)\frac{A_{\Delta}(\omega)}{B_{\Delta}(\omega)}. \label{eq:20}
\end{align}
It follows immediately from the relations \eqref{eq:19a} and \eqref{eq:19b} that the ratio \eqref{eq:20} satisfies the recurrence relations
\begin{align}
G_{\Delta}(\omega)
&= 	\frac{-1+\Delta \pm i\omega}{-\Delta \pm i\omega}G_{\Delta}(\omega \pm i). \label{eq:21}
\end{align}
We emphasize that, in contrast to the Euclidean case \cite{Ohya:2013xva}, $\omega$-dependence of $G_{\Delta}(\omega)$ is not uniquely determined from Eq.~\eqref{eq:21} and there are a number of nontrivial solutions that satisfy the recurrence relations.
Among them are the following ``minimal'' solutions:\footnote{Here ``minimal'' means that $G_{\Delta}(\omega)$ does not contain a factor $f(\omega) = f(\omega \pm i)$ such as $\mathrm{e}^{\pm2n\pi\omega}$ ($n\in\mathbb{Z}$).}
\begin{subequations}
\begin{align}
G^{\pm}_{\Delta}(\omega)
&\propto 	\mathrm{e}^{\pm\pi\omega}\Gamma(\Delta-i\omega)\Gamma(\Delta+i\omega), \label{eq:22a}\\
G^{A/R}_{\Delta}(\omega)
&\propto 	\frac{\Gamma(\Delta \pm i\omega)}{\Gamma(1-\Delta \pm i\omega)}, \label{eq:22b}
\end{align}
\end{subequations}
where proportional coefficients are $\omega$-independent.
The solutions $G^{\pm}_{\Delta}(\omega)$ correspond to the Fourier transforms of positive- and negative-frequency two-point Wightman functions, $G_{\Delta}^{+}(\omega) = \int_{-\infty}^{\infty}dt$ $\mathrm{e}^{i\omega t}\langle\mathcal{O}_{\Delta}(t)\mathcal{O}_{\Delta}(0)\rangle$ and $G_{\Delta}^{-}(\omega) = \int_{-\infty}^{\infty}dt\,\mathrm{e}^{i\omega t}\langle\mathcal{O}_{\Delta}(0)\mathcal{O}_{\Delta}(t)\rangle$,\footnote{$\langle\mathcal{O}_{\Delta}(t_{1})\mathcal{O}_{\Delta}(t_{2})\rangle \propto \left[\frac{\pi T}{\sinh(\pi T(t_{1}-t_{2}-i\epsilon))}\right]^{2\Delta}$.} both of which have simple poles at $\omega = \pm i(\Delta+n)$ ($n = 0,1,2,\cdots$).
On the other hand, the solutions $G^{A/R}_{\Delta}(\omega)$ correspond to the Fourier transforms of advanced and retarded two-point functions, $G_{\Delta}^{A}(\omega) = i\int_{-\infty}^{\infty}dt\,\theta(-t)\mathrm{e}^{i\omega t}\langle[\mathcal{O}_{\Delta}(t),\mathcal{O}_{\Delta}(0)]\rangle$ and $G_{\Delta}^{R}(\omega) = -i\int_{-\infty}^{\infty}dt\,\theta(t)\mathrm{e}^{i\omega t}\langle[\mathcal{O}_{\Delta}(t),\mathcal{O}_{\Delta}(0)]\rangle$, each of which has simple poles only on the positive and negative imaginary $\omega$-axis, $\omega = i(\Delta+n)$ and $\omega = -i(\Delta+n)$ ($n = 0,1,2,\cdots$), respectively.
It should be noted that the Fourier transforms of other real-time two-point functions such as the Pauli-Jordan commutator function $\langle[\mathcal{O}_{\Delta}(t),\mathcal{O}_{\Delta}(0)]\rangle$, the two-point Hadamard function $\langle\{\mathcal{O}_{\Delta}(t),\mathcal{O}_{\Delta}(0)\}\rangle$ and the Feynman propagator $\theta(t)\langle\mathcal{O}_{\Delta}(t)\mathcal{O}_{\Delta}(0)\rangle + \theta(-t)\langle\mathcal{O}_{\Delta}(0)\mathcal{O}_{\Delta}(t)\rangle$ are all given by appropriate linear combinations of \eqref{eq:22a} and \eqref{eq:22b}.

\subsection*{Acknowledgements}
The author is supported in part by JSPS Grand-in-Aid for Research Activity Startup \#15H06641.

\bibliographystyle{utphys}
\bibliography{Bibliography}

\providecommand{\href}[2]{#2}\begingroup\raggedright\begin{thebibliography}{1}

\bibitem{Ohya:2013xva}
S.~Ohya, ``{Recurrence relations for finite-temperature correlators via
  AdS$_{2}$/CFT$_{1}$},'' \href{http://dx.doi.org/10.1007/JHEP12(2013)011}{{\em
  JHEP} {\bf 12} (2013)  011},
\href{http://arxiv.org/abs/1309.2939}{{\tt arXiv:1309.2939 [hep-th]}}.

\bibitem{Ohya:2013vba}
S.~Ohya, ``{A Simple Derivation of Finite-Temperature CFT Correlators from the
  BTZ Black Hole},'' \href{http://dx.doi.org/10.14311/AP.2014.54.0142}{{\em
  Acta Polytech.} {\bf 54} (2014) 142--148},
\href{http://arxiv.org/abs/1312.7348}{{\tt arXiv:1312.7348 [hep-th]}}.

\bibitem{Spradlin:1999bn}
M.~Spradlin and A.~Strominger, ``{Vacuum states for $AdS_2$ black holes},''
  \href{http://dx.doi.org/10.1088/1126-6708/1999/11/021}{{\em JHEP} {\bf 11}
  (1999)  021},
\href{http://arxiv.org/abs/hep-th/9904143}{{\tt arXiv:hep-th/9904143
  [hep-th]}}.

\bibitem{Czech:2012be}
B.~Czech, J.~L. Karczmarek, F.~Nogueira, and M.~Van~Raamsdonk, ``{Rindler
  quantum gravity},''
  \href{http://dx.doi.org/10.1088/0264-9381/29/23/235025}{{\em Class. Quant.
  Grav.} {\bf 29} (2012)  235025},
\href{http://arxiv.org/abs/1206.1323}{{\tt arXiv:1206.1323 [hep-th]}}.

\bibitem{Parikh:2012kg}
M.~Parikh and P.~Samantray, ``{Rindler-AdS/CFT},''
\href{http://arxiv.org/abs/1211.7370}{{\tt arXiv:1211.7370 [hep-th]}}.

\bibitem{Iqbal:2008by}
N.~Iqbal and H.~Liu, ``{Universality of the hydrodynamic limit in AdS/CFT and
  the membrane paradigm},''
  \href{http://dx.doi.org/10.1103/PhysRevD.79.025023}{{\em Phys. Rev.} {\bf
  D79} (2009)  025023},
\href{http://arxiv.org/abs/0809.3808}{{\tt arXiv:0809.3808 [hep-th]}}.

\bibitem{Iqbal:2009fd}
N.~Iqbal and H.~Liu, ``{Real-time response in AdS/CFT with application to
  spinors},'' \href{http://dx.doi.org/10.1002/prop.200900057}{{\em Fortsch.
  Phys.} {\bf 57} (2009)  367--384},
\href{http://arxiv.org/abs/0903.2596}{{\tt arXiv:0903.2596 [hep-th]}}.

\end{thebibliography}\endgroup

\end{document}